\documentclass[12pt]{article}

\pdfoutput=1
\usepackage[totalheight = 23cm, totalwidth = 17cm]{geometry}
\usepackage{amssymb,amsmath,graphicx,subfigure,color,cite,bm}
\usepackage[colorlinks=true,linkcolor=blue,citecolor=blue]{hyperref}

\usepackage{ulem}

\newcommand{\pbh}{{\rm PBH}}
\newcommand{\calO}{{\cal O}}

\begin{document}

\begin{titlepage}

\begin{flushright}
APCTP-Pre2017-005
\end{flushright}

\vskip 2cm

\begin{center}

{\LARGE \bf
Small-scale structure and 21cm fluctuations \\ by primordial black holes \\
}

\vspace{1cm}

Jinn-Ouk Gong$^{\,a,b}$
and
Naoya Kitajima$^{\,a,c}$ \\

\vskip 1.0cm

{\it
$^a$Asia Pacific Center for Theoretical Physics, Pohang 37673, Korea
\\[2mm]
$^b$Department of Physics, POSTECH, Pohang 37673, Korea
\\[2mm]
$^c$Department of Physics, Nagoya University, Nagoya 464-8602, Japan
}

\vskip 1.0cm

\begin{abstract}

We discuss early structure formation of small scales sourced by primordial black holes (PBHs) which constitute a small part of present cold dark matter component. We calculate the mass function and power spectrum of haloes originated from the Poisson fluctuations of PBH number and show that the number of small haloes is significantly modified in the presence of PBHs even if their fraction accounts for only $10^{-4}$--$10^{-3}$ of total dark matter abundance. We then compute the subsequent 21cm signature from those haloes. We find that PBHs can provide major contributions at high redshifts within the detectability of future experiments such as Square Kilometer Array, and provide a forecast constraint on the PBH fraction.

\end{abstract}

\end{center}

\end{titlepage}

\newpage

\section{Introduction} 
\label{sec:intro}

Recently, LIGO observed gravitational wave signals produced by binary black hole mergers at high redshift, $z = 0.09$ \cite{Abbott:2016blz,Abbott:2016nmj}. It was found from GW150914 event that each black hole constituting the binary has a high-stellar mass, $M_{\rm BH} \sim 30 M_\odot$ and likewise from GW151226 $M_{\rm BH} \sim 10 M_\odot$. It is however difficult to explain the existence of such a system within the framework of conventional astrophysical processes \cite{TheLIGOScientific:2016htt,OLeary:2016ayz}. Among various challenges to explain the event \cite{Kinugawa:2014zha,Kinugawa:2015nla,Hartwig:2016nde}, primordial black hole (PBH) is one possible solution as pointed out in \cite{Bird:2016dcv,Clesse:2016vqa,Sasaki:2016jop} in such a way that two nearby PBHs can form a binary and eventually coalesce within the present age of the universe \cite{Nakamura:1997sm}. The estimated merger rate 9--240~Gpc$^{-3}$ yr$^{-1}$~\cite{TheLIGOScientific:2016pea} is consistent with the theoretical prediction.

PBH formation occurs through direct collapse of high density region in the early universe \cite{ZeldovichPBH} (see also \cite{Hawking:1971ei,Carr:1974nx} for other early works) seeded by the primordial fluctuations during inflation or subsequent cosmological phase transitions. The PBH mass is estimated as the energy of cosmic fluid stored in a sphere with the Hubble horizon radius at the formation time:
\begin{equation}
M_\pbh = \frac{4\pi}{3} H_{\rm f}^{-3} \rho_{r,{\rm f}} 
\approx 0.05 M_\odot \left( \frac{g_{*{\rm f}}}{100} \right)^{-1/2} \left( \frac{T_{\rm f}}{1~{\rm GeV}} \right)^{-2},
\end{equation}
where the subscript f represents that the quantity is evaluated at the PBH formation time, $\rho_r$ is the radiation energy density, $g_*$ is the number of the relativistic degrees of freedom and $M_\odot = 1.99\times10^{33}$ g is the solar mass. The present PBH density parameter can be calculated via 
\begin{equation}
\Omega_{\rm PBH0} h^2 = \frac{\rho_{\rm PBH, eq}}{\rho_{r,{\rm eq}}} \Omega_{m0}h^2 \approx 7 \times 10^9 \frac{\beta}{1-\beta} \left( \frac{M_\odot}{M_{\rm PBH}} \right)^{1/2} \, ,
\end{equation}
where the subscript eq represents the value at the matter-radiation equality, $\Omega_{m0} h^2 = 0.14$ is the present density parameter of matter content, and $\beta \equiv \rho_{\rm PBH,f}/\rho_{\rm f}$ is the density fraction of PBHs at the moment of formation. PBHs behave as a pressureless matter component in cosmic fluid and long-lived PBHs whose mass is larger than $10^{15}$ g can take a role of the present dark matter content (for recent reviews, see \cite{Carr:2009jm,Carr:2016drx}).

The existence of PBHs can be examined by various observations over wide mass range~\cite{Carr:2009jm,Carr:2016drx}. In particular, it is severely constrained by gravitational microlensing surveys \cite{Tisserand:2006zx,Niikura:2017zjd} and gravitational wave detection by pulsar timing observations \cite{Saito:2008jc,Saito:2009jt,Inomata:2016rbd,Orlofsky:2016vbd}. In a mass range of our interest $M_{\rm PBH} \sim 10 M_\odot$, there is a strong constraint from cosmic microwave background (CMB) observations \cite{Ricotti:2007au}, and the current density fraction of PBHs to total dark matter abundance is constrained as $f_\pbh \equiv \Omega_{\pbh0}/\Omega_{c0} \lesssim 10^{-4}\text{--}10^{-3}$ depending on the duty cycle parameter, where $\Omega_{c0}$ is the present cold dark matter density parameter. Interestingly, it is pointed out in \cite{Sasaki:2016jop} that the constraint is marginally satisfied if one consider the expected merger rate of such black holes from GW150914 event and the possibility can be tested in not-so-far future by the precise observation of CMB spectral distortion such as PIXIE \cite{Kogut:2011xw}.

Specific scenarios of the inflationary universe predict blue spectrum of the curvature perturbation and naturally lead to the formation of a significant number of PBHs after inflation \cite{Kawasaki:1997ju,Yokoyama:1998pt,Frampton:2010sw,Kawasaki:2016pql,Garcia-Bellido:2016dkw}. Non-thermal evolution of the universe during the post-inflationary epoch may well give rise to enhanced density perturbations on small scales that lead to the formation of PBHs consistent with the current observations \cite{Kawasaki:2012wr,Kohri:2012yw,Georg:2017mqk}. Note that the possibility of PBHs as the main component of dark matter is not completely ruled out if one considers a scenario predicting an extended mass spectrum~\cite{Inomata:2017okj,Kuhnel:2017rp}.

Because the PBH formation is a rare event, and the spatial distribution of PBHs is quite discrete, there is a Poisson fluctuation in the PBH number density as first pointed out in~\cite{Meszaros:1975ef} (see also \cite{Carr1977,Carr1983,Freese1983} for early works). Such a Poisson noise eventually behaves like an isocurvature mode of matter density perturbations on small scales. Since the power spectrum of Poisson noise is inversely proportional to the number density of PBH (see Section~\ref{subsec:powerspectra}), such a mode becomes more significant for heavier PBHs. Hence, the abundance of such PBHs can be constrained by observation of the Ly$\alpha$ forest~\cite{Afshordi:2003zb} and it can imprint non-trivial effects on the cosmic infrared background anisotropies~\cite{Kashlinsky:2016sdv}.

The existence of PBHs can affect the redshifted 21cm line fluctuations. For heavy, non-evaporating PBHs with $10M_\odot \lesssim M_{\rm PBH} \lesssim 10^8M_\odot$, X-ray photons are emitted due to matter accretion onto PBHs and nearby intergalactic medium (IGM) is heated and ionized. It can significantly alter the 21cm signature even though PBH accounts for only a tiny fraction of cold dark matter \cite{Tashiro:2012qe}. Light PBHs which evaporate in the dark age with $2\times10^{-20}M_\odot \lesssim M_{\rm PBH} \lesssim 10^{-16}M_\odot$ can also affect the 21cm brightness temperature through the heating of IGM due to the Hawking radiation \cite{Mack:2008nv}. More generally, the 21cm fluctuation is also sensitive to the number of minihaloes because the gas temperature and the density of neutral hydrogen can be significantly higher than the background values in such collapsed objects, leading to the enhancement of hyperfine transition of neutral hydrogen atom \cite{Iliev:2002gj,Furlanetto:2002ng}. Potentially, it has the detectability of small non-Gaussianity \cite{Chongchitnan:2012we} and the small-scale power spectrum altered by e.g. the neutrino mass, running spectral index for the curvature perturbation and warm dark matter \cite{Shimabukuro:2014ava}. The primordial blue-tilted isocurvature perturbation can also produce a number of minihaloes on small scales and affects the 21cm fluctuations \cite{Takeuchi:2013hza,Sekiguchi:2013lma}.

In this article, we calculate explicitly the halo mass function and halo power spectrum sourced by the Poisson fluctuations of PBHs. We show that the number of light haloes can significantly increase in the presence of such isocurvature mode sourced by PBHs. Further, we compute the resulting 21cm emission at high redshifts and find that those haloes provide dominant contributions which can be detected by future experiments such as Square Kilometer Array (SKA) \cite{SKA}, and that thus PBHs can be more strongly constrained by 21cm observations than the current constraint in the mass range $M_\odot \lesssim M_\pbh \lesssim 100 M_\odot$. This article is organized as follows. In Section~\ref{sec:halo}, we show the matter power spectrum and halo mass function from the PBH density fluctuation.  In Section~\ref{sec:21cm_halo} we compute the 21cm signature from haloes originated from PBHs. We summarize shortly the cosmological and astrophysical implication of our results in Section~\ref{sec:conc}.

\section{Minihalo formation with PBHs} 
\label{sec:halo}

The gravitational wave signals detected by LIGO indicate the existence of high-stellar mass black holes with $M_{\rm PBH} = \calO(10)M_{\odot}$. If these black holes are PBHs, there is an isocurvature mode on small scales due to the Poisson fluctuations in addition to the adiabatic Gaussian density perturbation. Such an isocurvature mode can alter the halo mass function on small scales as we will see in this section.

\subsection{Matter and PBH power spectra}
\label{subsec:powerspectra}

Let us start with the standard matter power spectrum of the density perturbation sourced by primordial adiabatic fluctuations. The density contrast in the position space is denoted as $\delta(\bm{x}) = \delta\rho(\bm{x})/\bar{\rho}$ where barred quantity expresses the background value and a smoothed density contrast over a volume $V$ is written as
\begin{equation}
\delta_{\rm s} = \int_V d^3x' \delta(\bm{x}') W(|\bm{x}-\bm{x}'|) = \int \frac{d^3k}{(2\pi)^3} e^{i \bm{k}\cdot\bm{x}} W(kR) \delta(\bm{k}) \, ,
\end{equation}
where $\delta(\bm{k})$ is the Fourier transformation of $\delta(\bm{x})$ and $W(kR)$ is the window function in the Fourier space corresponding to the smoothing scale $R$, with $W(\bm{x})$ being the one in the real space.
For the top-hat window function $W(x) = 3\Theta(R-x)/(4\pi R^3)$, it is given by
\begin{equation}
W(kR) = \frac{j_1(kR)}{kR} = \frac{3\sin(kR) - 3kR \cos(kR)}{(kR)^3} \, .
\end{equation}
The variance $\sigma^2(R)$ smoothed over the sphere with radius $R$ is calculated as
\begin{equation}
\label{eq:smoothedvariance}
\sigma^2(R) = \int^{\infty}_0 \frac{dk}{k} \frac{k^3 P(k)}{2 \pi^2} W^2(kR) \, ,
\end{equation}
where the power spectrum $P(k)$ is defined via
\begin{equation}
\label{eq:P}
\langle \delta(\bm{k}) \delta(\bm{k}') \rangle \equiv (2\pi)^2 \delta^{(3)}(\bm{k}+\bm{k}') P(k) \, ,
\end{equation}
with the brackets representing an ensemble average.

A horizon-sized region where the smoothed density contrast is larger than a threshold value collapses into a black hole at the time when the relevant scale reenters the horizon in the radiation dominated universe. Such an event is rare and thus PBHs are sparsely distributed in space. Thus, the number of PBHs, $N_\pbh$, follows the Poisson distribution function:
\begin{equation}
\label{eq:Poisson}
{\mathbb{P}}(N_\pbh) = \frac{\lambda^{N_\pbh} e^{-\lambda}}{N_\pbh!} \, ,
\end{equation}
where $\lambda$ is the mean and also the variance of $N_\pbh$, $\lambda = \langle N_\pbh \rangle = \langle \delta N_\pbh^2 \rangle$. The density contrast of PBH coming from the Poisson noise can be expressed by $N_\pbh = \bar{N}_\pbh (1+\delta_{\rm PBH})$. Note that $\delta_{\rm PBH}$ is independent of the adiabatic primordial density perturbation and thus behaves as an isocurvature mode. Replacing the ensemble average with the spatial average, the variance of the density contrast is related to its Fourier space average as
\begin{equation}
\langle \delta_{\rm PBH}^2 \rangle = \frac{1}{V} \int d^3x \delta_{\rm PBH}^2(\bm{x})  
= \frac{\langle |\delta_{\rm PBH}(\bm{k})|^2 \rangle}{V^2},
\end{equation}
where $V$ is the comoving volume of the universe. Then, the PBH power spectrum is simply, from the definition of the power spectrum \eqref{eq:P} and properties of the Poisson distribution \eqref{eq:Poisson},
\begin{equation}
\label{eq:PBH-P}
P_\pbh(k) = \frac{V}{\bar{N}_\pbh} = \frac{1}{{n}_{\rm PBH}} \, .
\end{equation}
Note that ${n}_{\rm PBH}$ is the {\it comoving} number density and remains constant under the cosmic expansion. Thus \eqref{eq:PBH-P} is fixed at the time of the PBH formation, which implies that if PBH mass function is monochromatic, \eqref{eq:PBH-P} is scale invariant. It is applicable for scales much larger than the Hubble horizon scale at the time of the PBH formation.

\begin{figure}[h]
\centering
\subfigure[$M_\pbh = M_\odot$]{
\includegraphics [width = 7.5cm, clip]{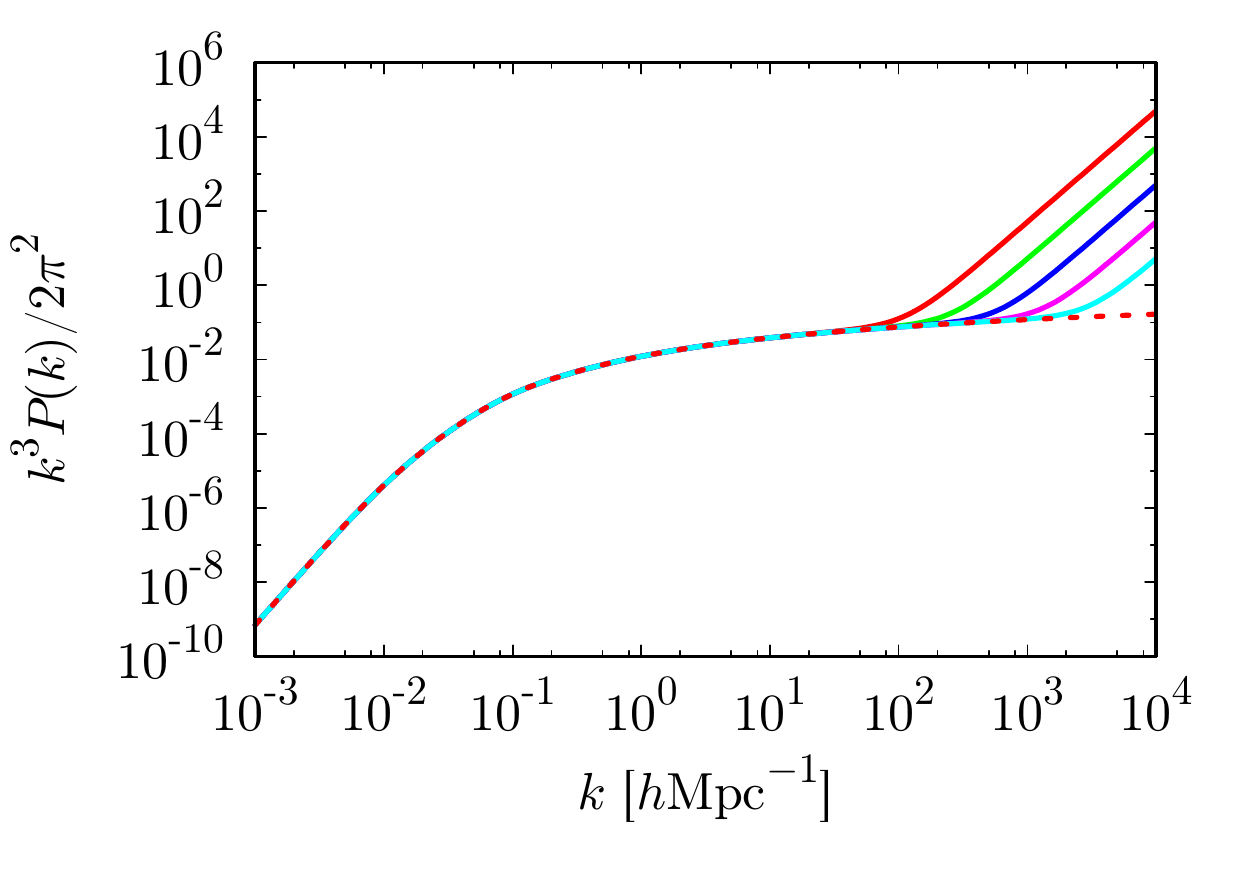}
\label{subfig:linearpower1}
}
\subfigure[$M_\pbh = 100M_\odot$]{
\includegraphics [width = 7.5cm, clip]{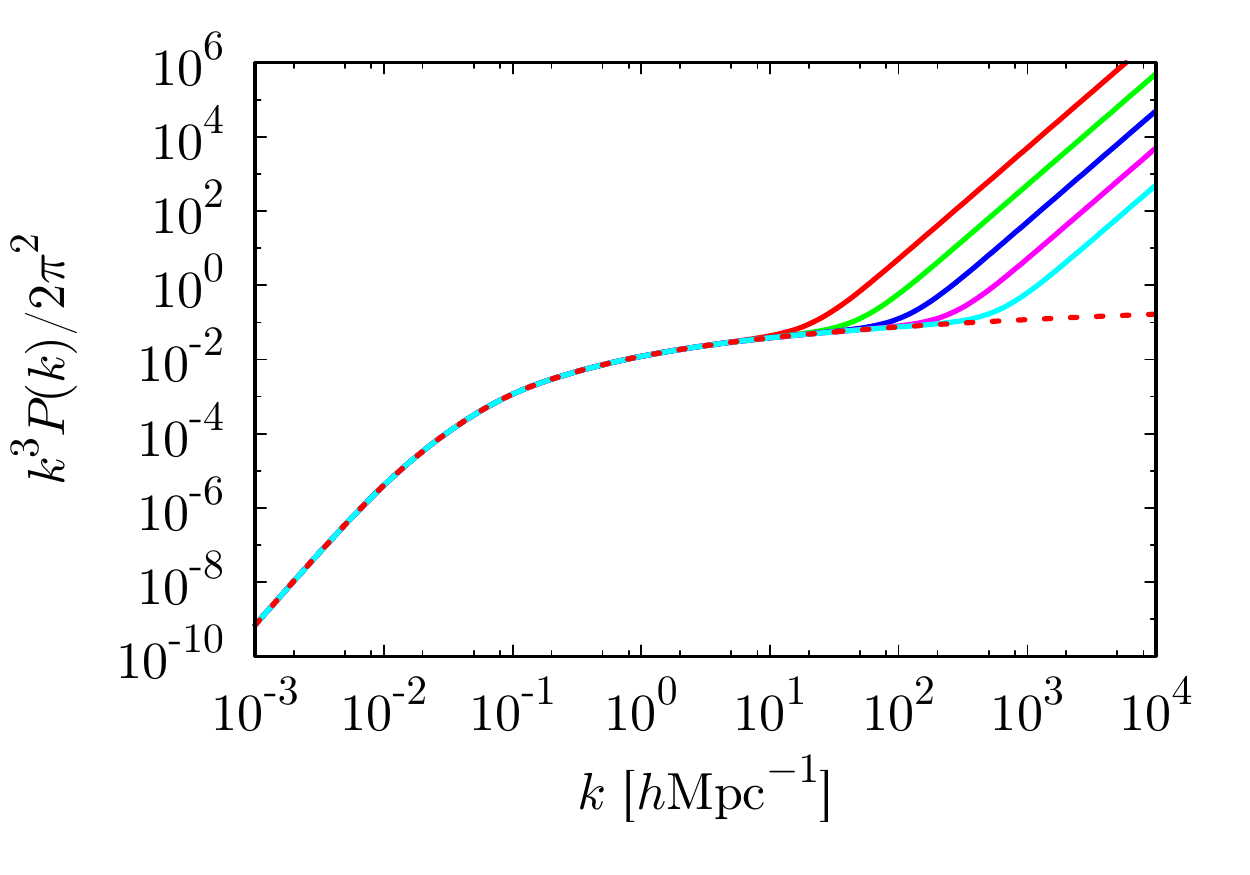}
\label{subfig:linearpower2}
}
\caption{
Linear matter power spectrum as the sum of the adiabatic contribution and Poisson fluctuation from PBHs at redshift $z=20$. We have set $M_{\rm PBH} = M_\odot$ ($100M_\odot$) in the left (right) panel and $f_{\rm PBH} = 10^{-1}$, $10^{-2}$, $10^{-3}$, $10^{-4}$, $10^{-5}$ from top to bottom in each panel. The red dotted lines represent conventional prediction without PBHs.
}
\label{fig:linearpower}
\end{figure}

Since $\delta_\pbh$ is an isocurvature mode, the transfer function can be approximately given by \cite{Peacock:1999ye}
\begin{equation}
T_{\rm iso}(k) = 
\begin{cases} 
\dfrac{3}{2}(1+z_{\rm eq}) ~~&\text{for}~~ k > k_{\rm eq} 
\\[2mm] 
0 ~~&\text{otherwise} 
\end{cases} 
\, ,
\end{equation}
where $1+z_{\rm eq} \simeq 3400$ is the redshift at the matter-radiation equality. This approximation is valid in our case because the isocurvature contribution is highly negligible on scales near and below $k_{\rm eq}$. In fact, we have checked that more precise fitting formula of the transfer function \cite{Bardeen:1985tr} does not change the subsequent results. Taking into account the linear growth factor $D(z)$ normalized by $D(0)=1$, the PBH power spectrum at redshift $z$ is given by\footnote{
Conventionally, the isocurvature perturbation is characterized as
\begin{equation*}
P_{\rm iso}(k) = \frac{2\pi^2}{k^3} A_{\rm iso} \bigg(\frac{k}{k_0}\bigg)^{n_{\rm iso}-1},
\end{equation*}
where $k_0$ is the pivot scale and $A_{\rm iso}$ and $n_{\rm iso}$ represent the amplitude and spectral index of the isocurvature perturbation at the pivot scale respectively. Our case corresponds to $A_{\rm iso} = 3.2\times 10^{-12} f_\pbh (M_\pbh/30M_\odot)$ or $A_{\rm iso}/A_{\rm adi} = 1.5 \times 10^{-3} f_\pbh (M_\pbh/30M_\odot)$ and $n_{\rm iso} = 4$ with $A_{\rm adi} = 2.2 \times 10^{-9}$ being the amplitude of the adiabatic curvature at the pivot scale.
}
\begin{equation}
\begin{split}
P_\pbh(k,z) &= \frac{9}{4} (1+z_{\rm eq})^2 f_{\rm PBH}^2 n_{\rm PBH}^{-1} D^2(z) 
\\
& \approx 2.5 \times 10^{-2} f_{\rm PBH} \bigg(\frac{M_{\rm PBH}}{30M_\odot} \bigg) D^2(z)~{\rm Mpc^3} ~~~\text{for}~~k>k_{\rm eq}\, .
\end{split}
\label{eq:pbh_powerspec}
\end{equation}
Since ${n}_{\rm PBH}$ is a conserved value, we have made use of the relation $M_{\rm PBH} n_{\rm PBH} = f_{\rm PBH} \Omega_{c0} \rho_{\rm cr0}$, where $\rho_{\rm cr0} = 2.78 \times 10^{11}h^2M_\odot/{\rm Mpc}^3$ is the present critical density. The total power spectrum is the sum of the contribution from the standard adiabatic perturbation and that from the Poisson fluctuation of PBHs. Figure~\ref{fig:linearpower} shows the total power spectrum in the presence of the PBH Poisson fluctuation.

\subsection{Halo mass function and power spectrum from PBHs}
\label{subsec:halomassfunc}

One can show that, by using the central limit theorem, the Poisson distribution for the PBH number can be translated into the Gaussian distribution for the density contrast with $\langle \delta_{\rm PBH} \rangle = 0$ and $\langle\delta_{\rm PBH}^2 \rangle = 1/\bar{N}_\pbh$ in the limit of $N_\pbh \gg 1$ and $\delta_{\rm PBH} \ll 1$. Here we assume the validity of this translation can be extrapolated to the region of $\delta_\pbh \gtrsim 1$ and make use of the standard Press-Schechter formalism~\cite{Press:1973iz} to count the halo number in the case where the Poisson fluctuation is included.

Let us consider a smoothed density field over a sphere with radius $R$. One can trade the smoothing radius $R$ in terms of the corresponding mass scale $M$ by inverting the halo mass $M = 4\pi\Omega_m\rho_{\rm cr}R^3/3$ within the comoving radius $R$ and find 
\begin{equation}
\frac{R}{h^{-1}{\rm Mpc}} = 0.95\times 10^{-4}~ \Omega_{m0}^{-1/3} \bigg( \frac{M}{h^{-1}M_\odot} \bigg)^{1/3} \, .
\end{equation}
The halo mass function denoted by $dn/dM$ is defined by the number of halos which have masses in a range between $M$ and $M+dM$, and conventionally it is expressed as
\begin{equation}
\frac{dn}{dM} = \frac{\rho_m}{M} \frac{d\log\sigma^{-1}}{dM} f(\sigma) \, ,
\end{equation}
where $f(\sigma)$ is a fitting function and we use the formula derived in~\cite{Sheth:1999mn}.

\begin{figure}[tp]
\centering
\subfigure[$M_{\rm PBH}=M_\odot$]{
\includegraphics [width = 7.5cm, clip]{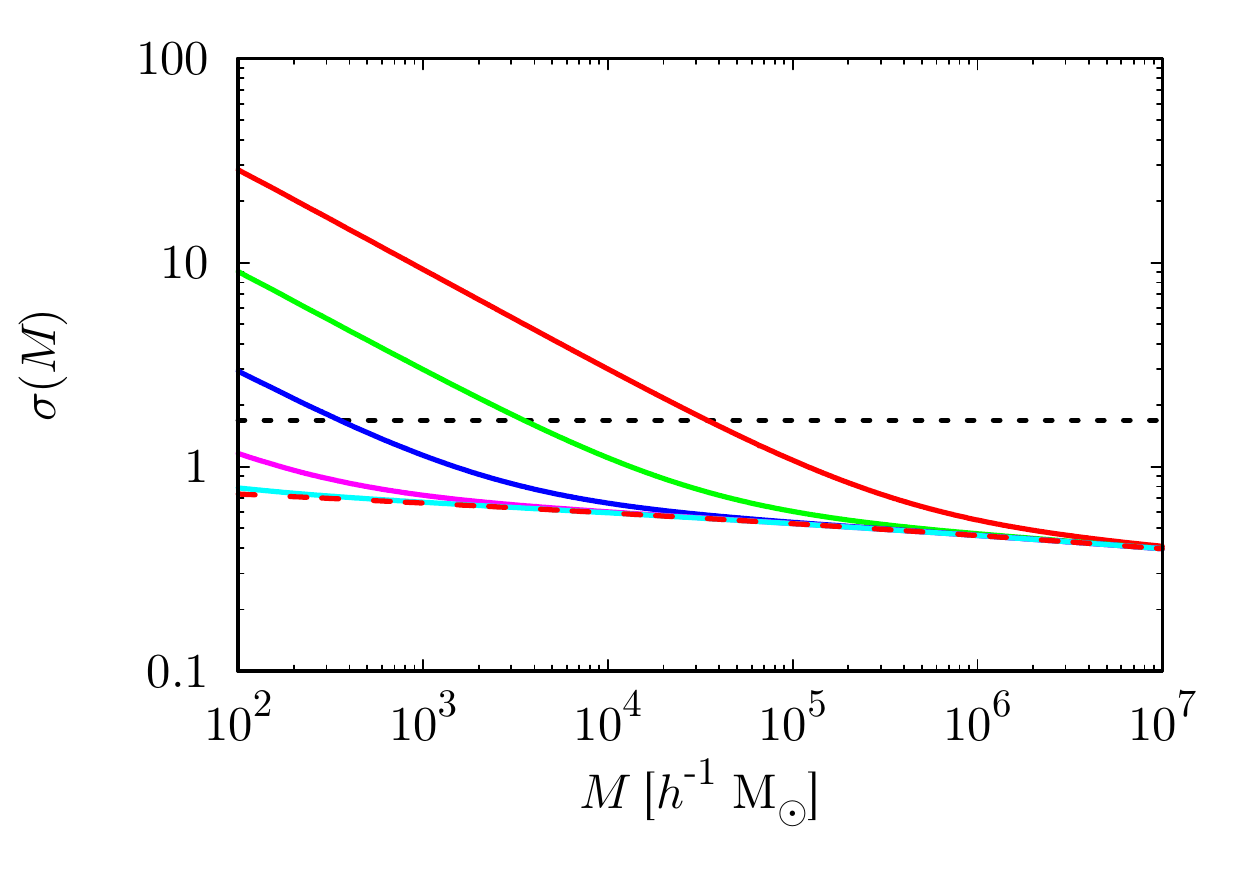}
\label{subfig:sigmaM1}
}
\subfigure[$M_{\rm PBH}=100M_\odot$]{
\includegraphics [width = 7.5cm, clip]{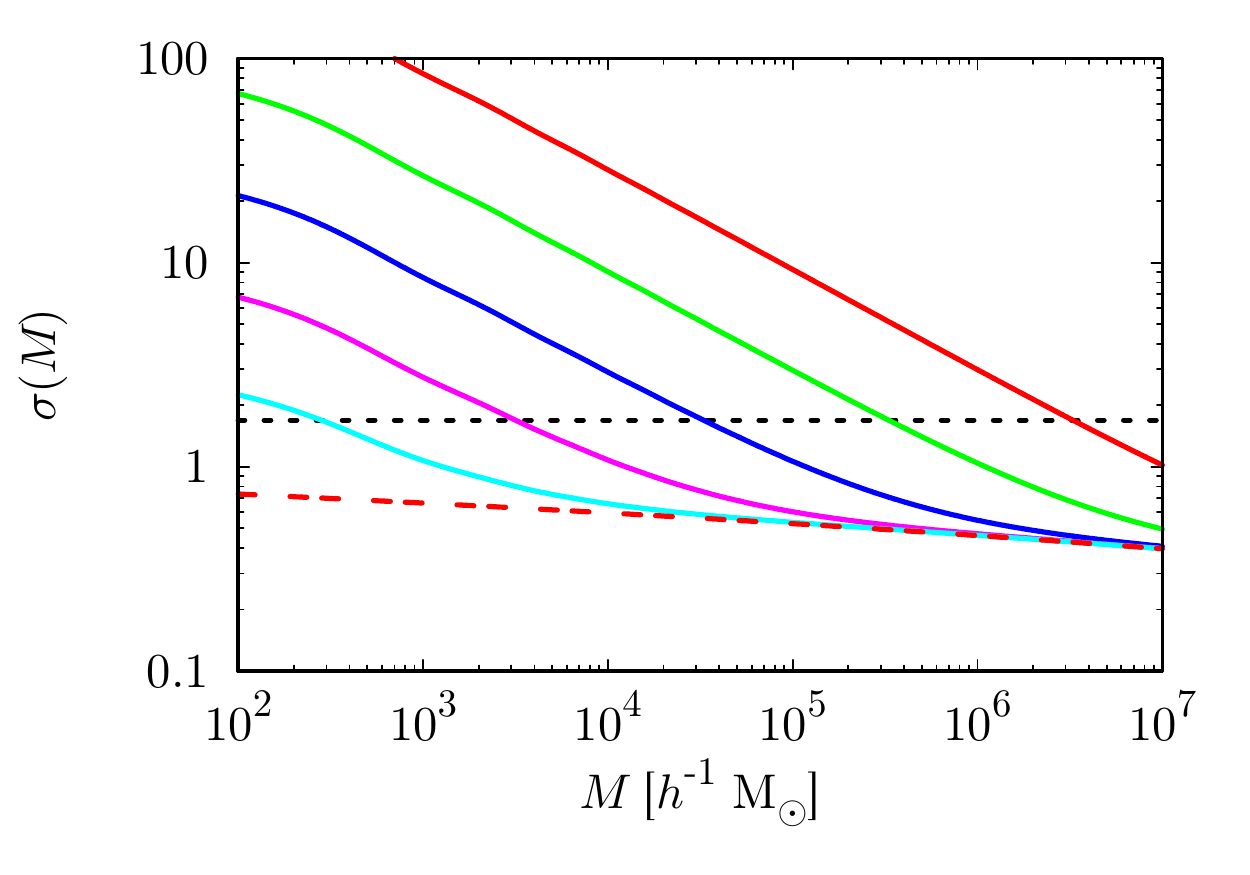}
\label{subfig:sigmaM2}
}
\caption{
Smoothed variance of linear density perturbation at redshift $z=20$. We have set $M_{\rm PBH} = M_\odot$ ($100M_\odot$) in the left (right) panel and $f_{\rm PBH} = 10^{-1}$, $10^{-2}$, $10^{-3}$, $10^{-4}$, $10^{-5}$ from top to bottom (solid lines). The dashed lines corresponds to the pure adiabatic case. The horizontal dotted black line shows the critical density contrast for spherical collapse, $\delta_c = 1.69$.
}
\label{fig:sigmaM}
\end{figure}

\begin{figure}[tp]
\centering
\subfigure[$M_{\rm PBH}=M_\odot$]{
\includegraphics [width = 7.5cm, clip]{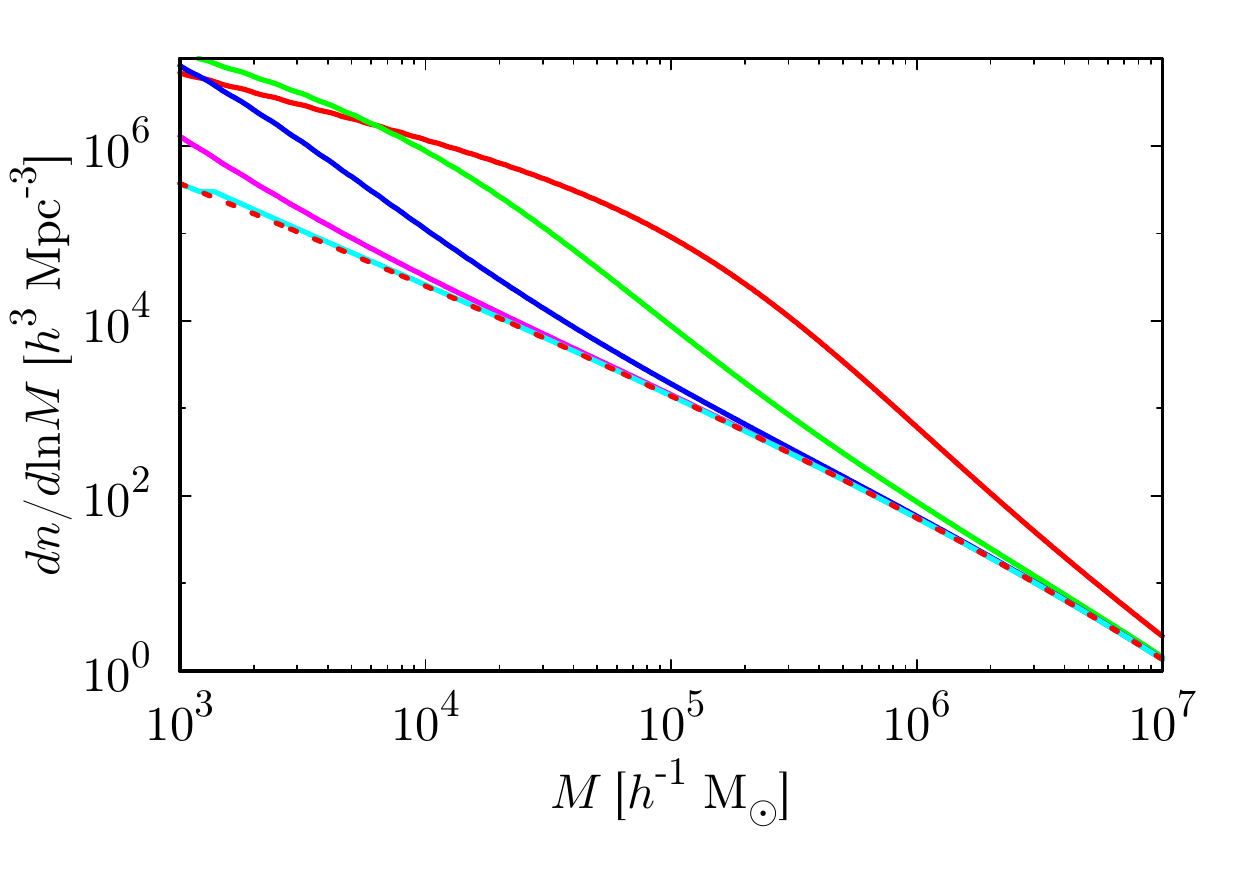}
\label{subfig:mf1a}
}
\subfigure[$M_{\rm PBH}=10M_\odot$]{
\includegraphics [width = 7.5cm, clip]{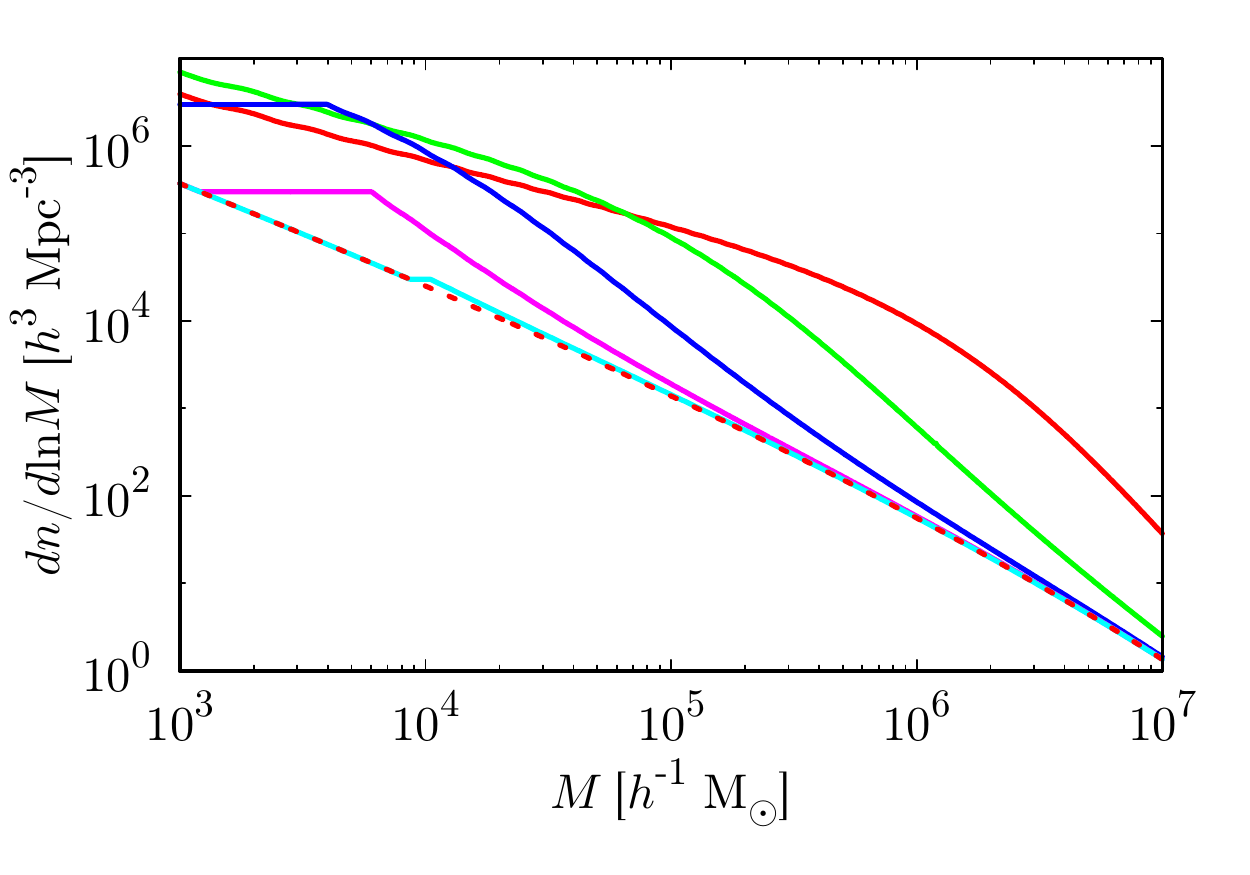}
\label{subfig:mf1b}
}
\\
\subfigure[$M_{\rm PBH}=100M_\odot$]{
\includegraphics [width = 7.5cm, clip]{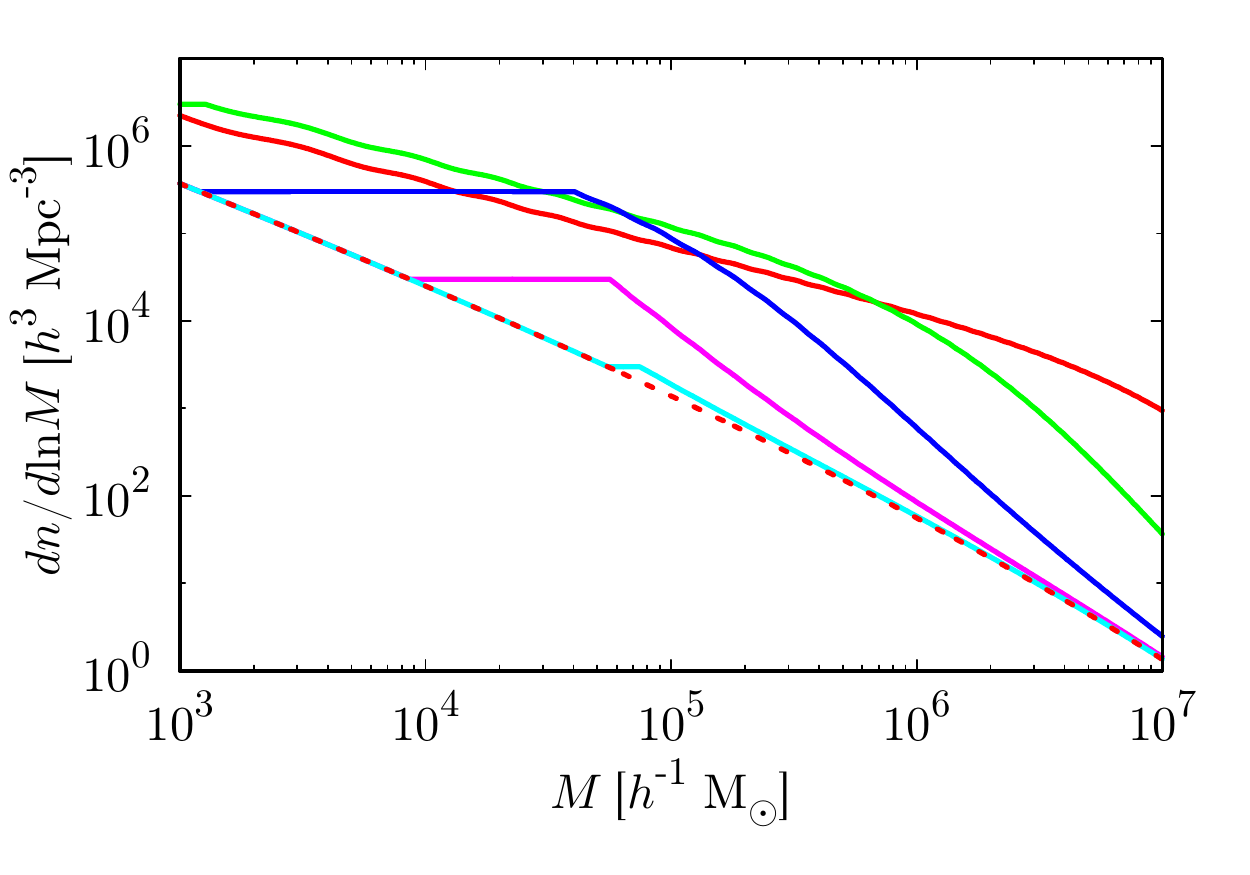}
\label{subfig:mf1c}
}
\subfigure[$M_{\rm PBH}=1000M_\odot$]{
\includegraphics [width = 7.5cm, clip]{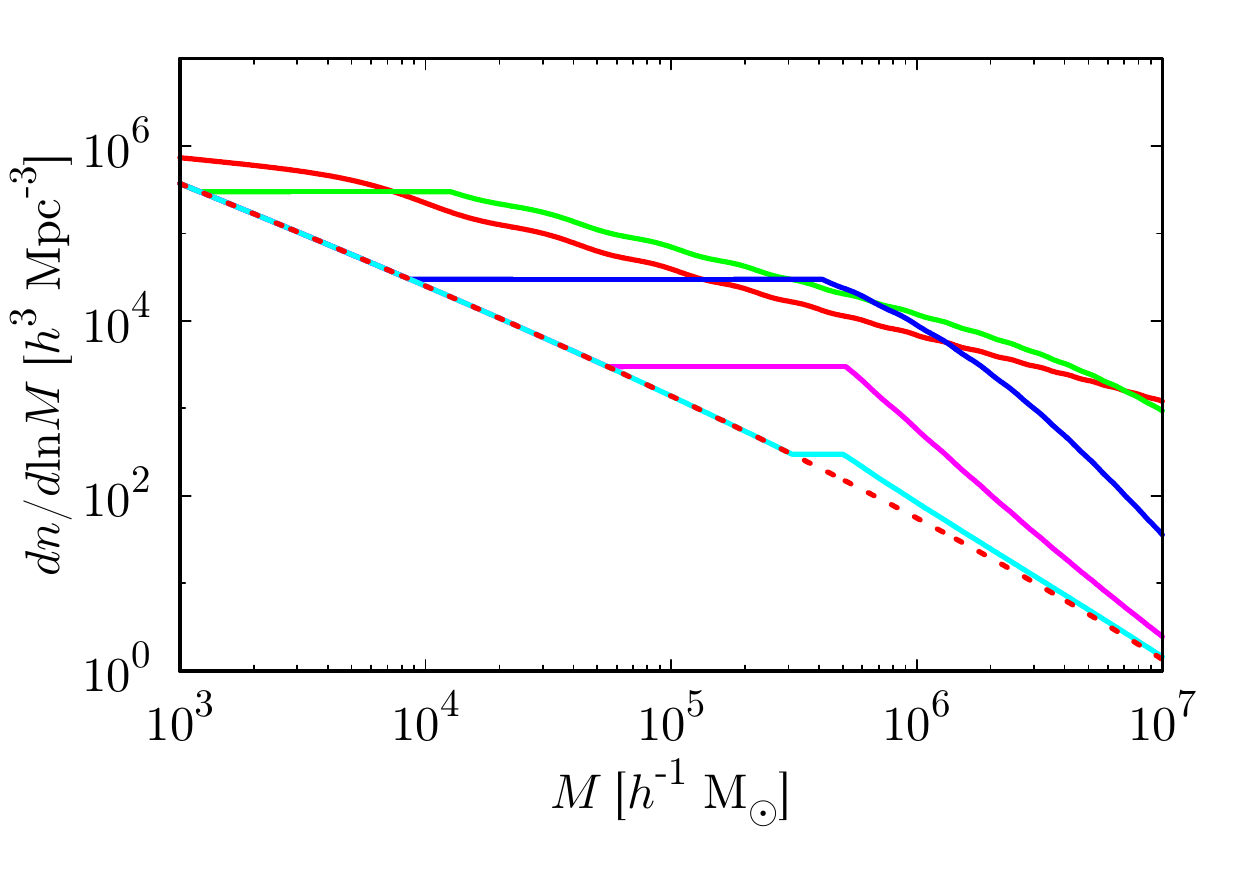}
\label{subfig:mf1d}
}
\caption{
The halo mass function with logarithmic mass interval at redshift $z=20$. We have set $M_{\rm PBH} = M_\odot$ (top left), $10M_\odot$ (top right), $100M_\odot$ (bottom left) and $1000M_\odot$ (bottom right). In each panel, $f_{\rm PBH} = 10^{-1}$, $10^{-2}$, $10^{-3}$, $10^{-4}$, $10^{-5}$ from top to bottom at the right endpoint.
}
\label{fig:mf1}
\end{figure}

PBHs contribute additional power \eqref{eq:pbh_powerspec} for the formation of dark matter haloes through the variance \eqref{eq:smoothedvariance}. Being constant on all scales, \eqref{eq:pbh_powerspec} is dominant on small scales or, equivalently, for haloes with small masses. Figure~\ref{fig:sigmaM} shows $\sigma(M)$ numerically computed by adding \eqref{eq:pbh_powerspec} to $P(k)$ in \eqref{eq:smoothedvariance}. Dashed lines are the standard adiabatic results, while solid lines exhibit an enhancement due to the Poisson fluctuations. Because $\sigma(M)$ from PBHs exceeds the critical density, even though PBHs account for only a small fraction of total dark matter, their Poisson fluctuations can seed small-scale structure at high redshift. We use this $\sigma(M)$ for the computation of the halo mass function accordingly. In addition, we postulate an upper bound on the halo mass function in the case where the Poisson fluctuation dominates the adiabatic one. This is based on the fact that the halo formation in that case occurs around at least a single PBH and thus the number of haloes cannot exceed that of PBHs. One can estimate the comoving number density of PBHs $n_\pbh$ as 
\begin{equation}
n_\pbh = \frac{\rho_{\rm PBH}}{M_{\rm PBH}} = \frac{f_{\rm PBH} \Omega_{c0} \rho_{\rm cr0}}{M_{\rm PBH}} \sim 10^5 {\rm Mpc}^{-3} \bigg(\frac{f_{\rm PBH}}{10^{-4}}\bigg) \bigg(\frac{30M_{\odot}}{M_{\rm PBH}} \bigg) \, .
\label{eq:PBH_number}
\end{equation}
Hence, we conservatively impose the maximum value of the halo mass function given by \eqref{eq:PBH_number} if the Poisson fluctuation is the dominant source of the halo formation. We set the low-mass cutoff as well to be the PBH mass in that case. The resulting mass function with logarithmic mass interval at redshift $z = 20$ is shown in Figure~\ref{fig:mf1}.

\begin{figure}[tp]
\centering
\subfigure[$M_\pbh = M_\odot$]{
\includegraphics [width = 7.5cm, clip]{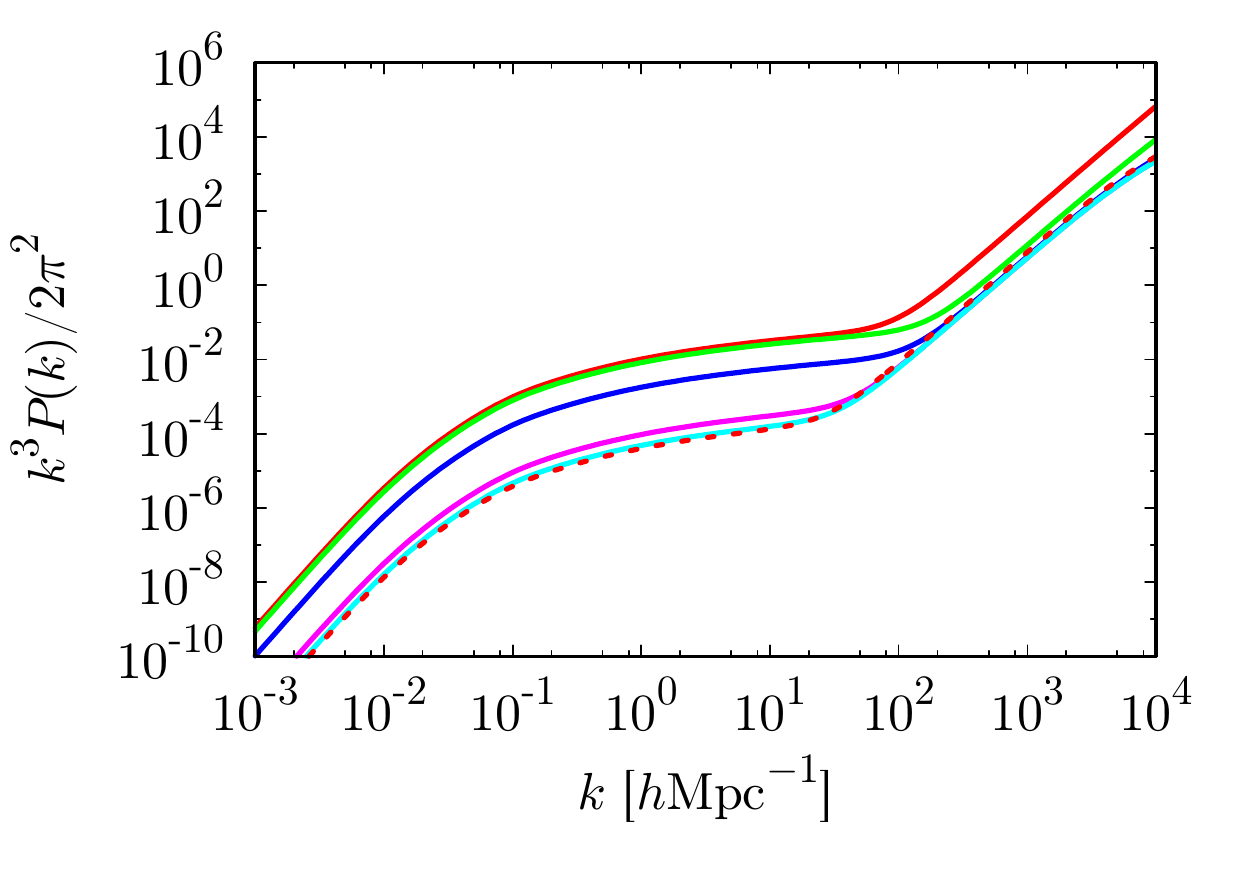}
\label{subfig:halopower1}
}
\subfigure[$M_\pbh = 100M_\odot$]{
\includegraphics [width = 7.5cm, clip]{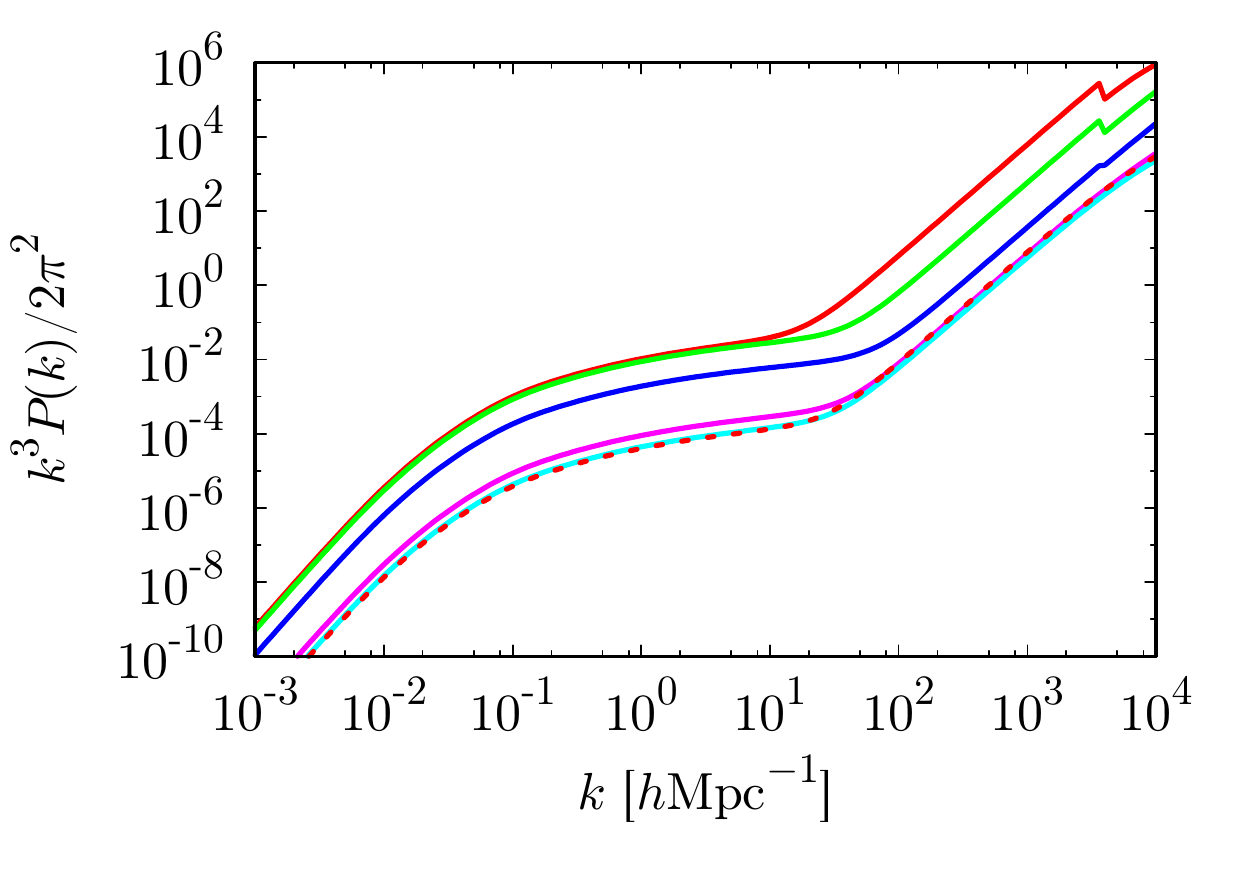}
\label{subfig:halopower2}
}
\caption{
Halo power spectrum based on \eqref{eq:Phalo}. We have set $M_{\rm PBH} = M_\odot$ ($100M_\odot$) in the left (right) panel and $f_{\rm PBH} = 10^{-1}$, $10^{-2}$, $10^{-3}$, $10^{-4}$, $10^{-5}$ from top to bottom in each panel. The dotted red lines show the prediction from the conventional adiabatic fluctuation without PBHs.
}
\label{fig:halopower}
\end{figure}

Finally, let us consider the matter power spectrum based on the halo model (for a review, see \cite{Cooray:2002dia}). In the halo model, the power spectrum is divided into two parts, one halo term and two halo term:
\begin{equation}
\label{eq:Phalo}
P_{\rm halo}(k) = P_{\rm 1h}(k) + P_{\rm 2h} (k) \, ,
\end{equation}
where 
\begin{align}
P_{\rm 1h}(k) & = \int dm \frac{dn}{dm} \bigg(\frac{m}{\bar{\rho}} \bigg)^2 |u(k,m)|^2 \, ,
\\[2mm]
P_{\rm 2h}(k) & = \int dm_1 \frac{dn}{dm_1} \bigg(\frac{m_1}{\bar{\rho}}\bigg) u(k,m_1) \int dm_2 \frac{dn}{dm_2} \bigg(\frac{m_2}{\bar{\rho}} \bigg) u(k,m_2) b(m_1) b(m_2) P(k) \, .
\end{align}
Here, $b(m)$ is the linear bias, $\bar{\rho} = \int dm m dn/dm$ is the mean matter density in haloes with $dn/dm$ as shown in Figure~\ref{fig:mf1}, and $u(k,m)$ is the normalized Fourier transformation of the density profile
\begin{equation}
u(k,m) \equiv \int^{r_{\rm vir}}_0 dr 4 \pi r^2 \frac{\sin kr}{kr} \frac{\rho_{\rm NFW}(r,m)}{m} \, ,
\end{equation}
where $r_{\rm vir}$ is the virial cutoff radius and we have adopted the Navarro-Frenk-White (NFW) density profile \cite{Navarro:1996gj}. Figure~\ref{fig:halopower} shows the resulting halo power spectrum under the presence of the PBH Poisson fluctuations. Since the halo formation is significantly promoted by the PBH Poisson fluctuations, the halo power spectrum is overall enhanced especially for larger $f_\pbh$ and $M_\pbh$. The discontinuities near the right endpoint in Figure~\ref{subfig:halopower2} come from the low-mass cutoff of the haloes from PBHs.

\section{21cm signature from minihaloes} 
\label{sec:21cm_halo}

In this section, we discuss the 21cm signature as a consequence of the minihalo formation from the PBH Poisson fluctuations and show the detectability and the forecast of the constraint on the PBH abundance by future observations such as SKA. Here we adopt the truncated isothermal sphere \cite{Shapiro:1998zp,Iliev:2001he} as a baryonic halo model relevant for the 21cm line emission/absorption and follow the analysis in \cite{Iliev:2002gj,Chongchitnan:2012we}.

The 21cm emission/absorption signal can be calculated in terms of the spin temperature, $T_S$, which is defined through the relative number density of the neutral hydrogen atom between the singlet ($n_0$) and triplet ($n_1$) hyperfine splitting states, 
\begin{equation}
\frac{n_1}{n_0} = 3\exp \bigg(-\frac{T_*}{T_S} \bigg) \, , 
\end{equation}
where $T_* = 0.068~{\rm mK}$ is the temperature corresponding to the energy splitting between two states. The spin temperature is formally expressed as 
\begin{equation}
T_S = \frac{T_{\rm CMB} + y_\alpha T_\alpha + y_c T_K}{1+ y_\alpha + y_c} \, ,
\end{equation}
where $T_{\rm CMB}$, $T_K$, $T_\alpha$, $y_\alpha$ and $y_c$ are respectively the CMB temperature, the kinetic temperature, the Ly$\alpha$ color temperature, the radiative coupling constant and the collisional coupling constant \cite{Madau:1996cs}. Here we assume that bright luminous sources for UV and X-ray photons are not yet abundant and Ly$\alpha$ pumping is not efficient. Hence we set $y_\alpha = 0$. For the case of H-H collision, $y_c$ is given by
\begin{equation}
y_c = \frac{T_* n_{\rm HI} \kappa}{A_{10}T_K} \, ,
\end{equation}
where $n_{\rm HI}$ is the number density of the neutral hydrogen, $A_{10} = 2.85\times 10^{-15}~{\rm s}^{-1}$ is the Einstein $A$ coefficient for the 21cm transition and $\kappa$ is approximately given by \cite{Kuhlen:2005cm}
\begin{equation}
\kappa = 3.1 \times 10^{-11}\bigg( \frac{T_K}{1~{\rm K}} \bigg)^{0.357} \exp(-32~{\rm K}/T_K)~{\rm cm^3~s^{-1}} \, .
\end{equation}

The effect of minihaloes or IGM on the 21cm radiation can be measured as the brightness temperature. By solving the equation of radiative transfer, the brightness temperature for photons with frequency $\nu$ coming through a single minihalo with impact parameter $\alpha$ can be expressed as
\begin{equation}
T_b(\nu,\alpha,z) = T_{\rm CMB}(z)e^{-\tau(\nu)} + \int^{\infty}_{-\infty}dR\,T_S(\ell) e^{-\tau(\nu,R)} \frac{\partial \tau}{\partial R} \, ,
\label{eq:T_b}
\end{equation}
where $T_{\rm CMB}(z) = 2.75{\rm K}(1+z)$ is the CMB temperature at redshift $z$ and $\ell$ is the radial comoving distance from the center of the minihalo satisfying $\ell^2 = R^2 + (\alpha r_t)^2$ with $r_t$ being a cutoff radius of the minihalo. The total optical depth for photons with frequency $\nu$ through a minihalo along the line of sight can be calculated as \cite{Furlanetto:2002ng} 
\begin{equation}
\tau (\nu) = \frac{3c^2 A_{10} T_*}{32 \pi \nu_*^2} \int^{\infty}_{-\infty}dR \frac{n_{\rm HI}(\ell) \phi(\nu,\ell)}{T_S(\ell)} \, ,
\end{equation}
where $\phi(\nu,\ell)$ is the line profile given by
\begin{equation}
\phi(\nu,\ell) = \frac{1}{\Delta\nu\sqrt{\pi}} \exp \bigg[ -\frac{(\nu - \nu_*)^2}{\Delta\nu^2} \bigg] 
\quad \text{where} \quad 
\Delta\nu = \nu_* \sqrt{\frac{2k_B T_K(\ell)}{m_H c^2}}
\end{equation}
for the Doppler broadening with $\nu_* = 1.42$~GHz being the rest-frame frequency corresponding to the 21cm radiation. $\tau(\nu,R)$ in the second term in \eqref{eq:T_b} is given by
\begin{equation}
\tau(\nu,R) = \tau_{\rm IGM} + \frac{3c^2A_{10}T_*}{32\pi\nu_*^2} \int^R_{-\infty}dR' \frac{n_{\rm HI}(\ell') \phi(\nu, \ell')}{T_S(\ell')} \, ,
\end{equation}
where $\tau_{\rm IGM}$ is the IGM optical depth
\begin{equation}
\tau_{\rm IGM}(z) = \frac{3c^3A_{10}T_*n_{\rm HI}(z)}{32\pi \nu_*^3 T_S(z) H(z)} \, .
\end{equation}

From (\ref{eq:T_b}), one obtains the differential brightness temperature of the 21cm radiation from minihaloes with respect to the background CMB temperature defined by
\begin{equation}
\delta T_b = \frac{\langle T_b \rangle}{1+z} - T_{\rm CMB}(0) \, ,
\end{equation}
where an angular bracket represents the average over the halo cross section, $\langle T_b \rangle = \int T_b(\alpha) dA/A$ with $A = \pi r_t^2$. Since there exist minihaloes with various masses at redshift $z$, the observed differential brightness temperature is averaged over halo mass and is written as
\begin{equation}
\overline{\delta T_b} = \frac{c(1+z)^4}{\nu_* H(z)} \int^{M_{\rm max}}_{M_{\rm min}} \Delta \nu_{\rm eff} \delta T_b (M) A \frac{dn}{dM} dM \, ,
\end{equation}
where $\Delta \nu_{\rm eff} = [\phi(\nu_*)(1+z)]^{-1}$ is the effective redshifted line width. $M_{\rm max}$ is determined by the virial temperature $T_{\rm vir} = 10^4$ K and $M_{\rm min} = \max(M_{\rm J},\,M_\pbh)$ with $M_{\rm J}$ being the Jeans mass given by
\begin{equation}
M_{\rm J}= 5.7 \times 10^3 M_\odot \bigg( \frac{1+z}{10} \bigg)^{3/2} \, .
\end{equation}
Finally, we obtain the $q$-$\sigma$ fluctuation of the differential brightness temperature smoothed over the survey volume with beam angle $\Delta\theta_{\rm beam}$ and frequency band width $\Delta\nu_{\rm band}$ \cite{Iliev:2002gj}, 
\begin{equation}
\label{eq:rms-Tb}
\langle \delta T_b^2 \rangle^{1/2} = q \sigma_p(\Delta\theta_{\rm beam},\Delta\nu_{\rm band}) \beta(z)\overline{\delta T}_b \, ,
\end{equation}
where $\sigma_p$ denotes the root-mean-square of the fluctuation of the linear density perturbation smoothed over the observational cylinder with comoving radius $R = \Delta\theta_{\rm beam}(1+z) D_A(z)/2$ and length $L = (1+z)c H(z)^{-1}(\Delta\nu_{\rm band}/\nu)$ with $D_A$ being the angular diameter distance \cite{Tozzi:1999zh}. $\beta(z)$ is the flux-weighted average of the bias $b(z,M)$ \cite{Chongchitnan:2012we}.

\begin{figure}[tp]
\centering
\subfigure[$M_{\rm PBH} = M_{\odot}$]{
\includegraphics [width = 7.5cm, clip]{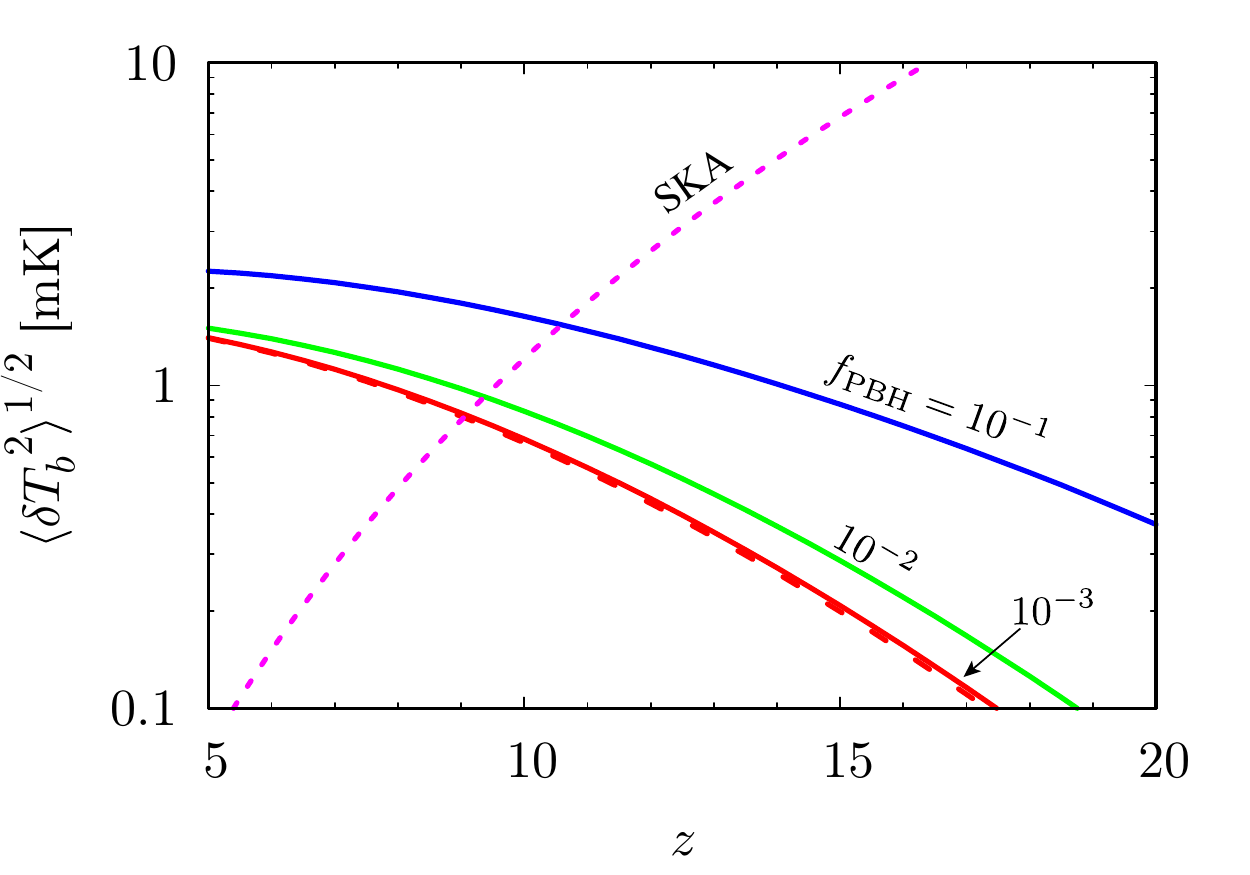}
\label{subfig:deltaT_b_1}
}
\subfigure[$M_{\rm PBH} = 10M_{\odot}$]{
\includegraphics [width = 7.5cm, clip]{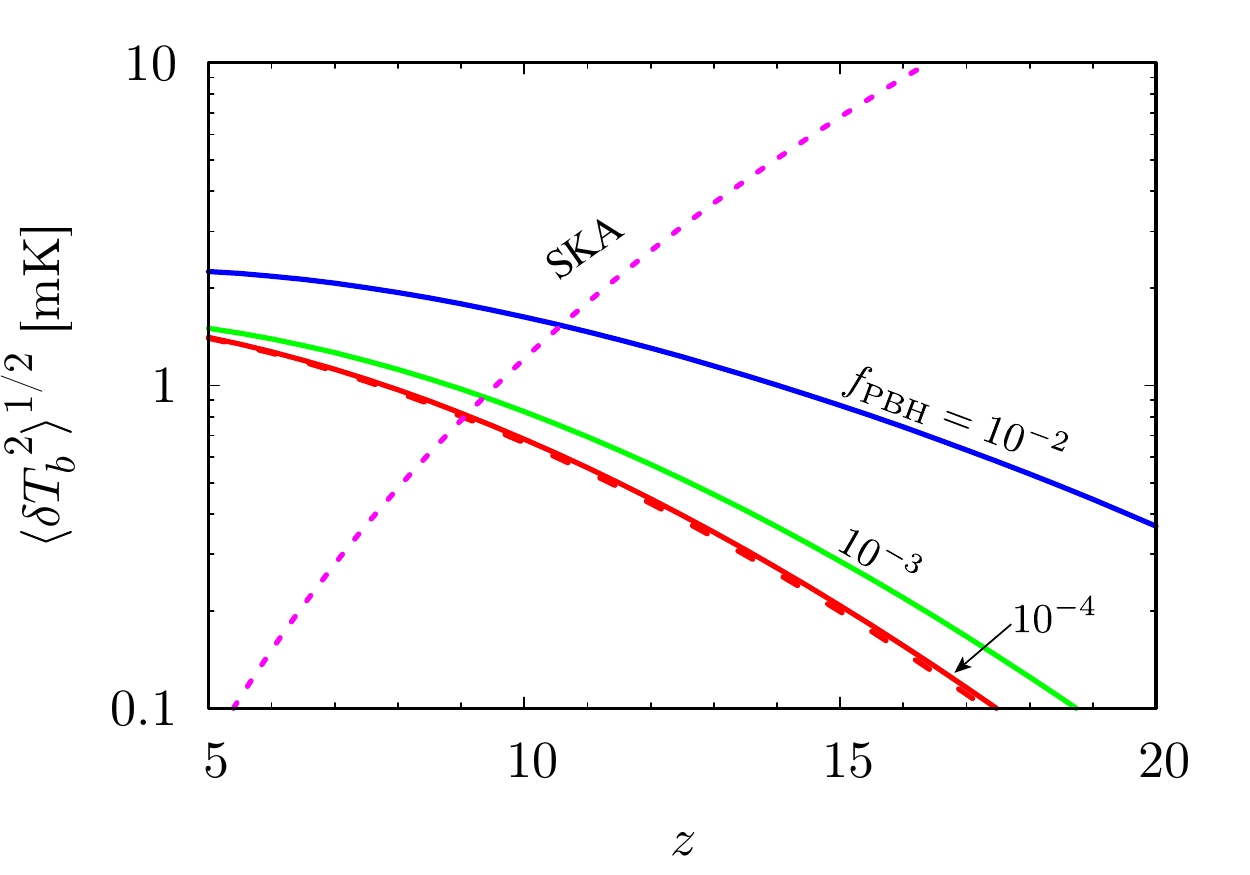}
\label{subfig:deltaT_b_2}
}
\subfigure[$M_{\rm PBH} = 100M_{\odot}$]{
\includegraphics [width = 7.5cm, clip]{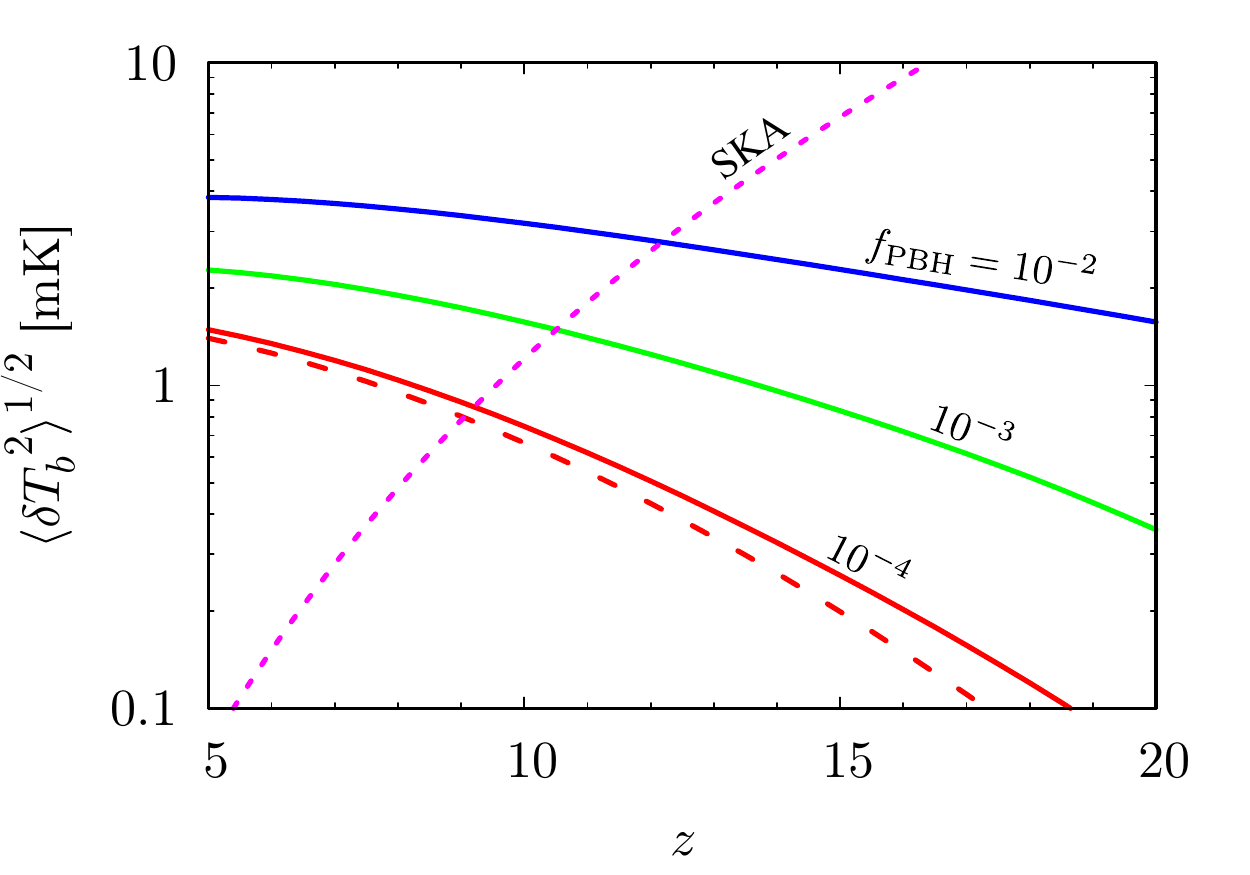}
\label{subfig:deltaT_b_3}
}
\subfigure[$M_{\rm PBH} = 1000M_{\odot}$]{
\includegraphics [width = 7.5cm, clip]{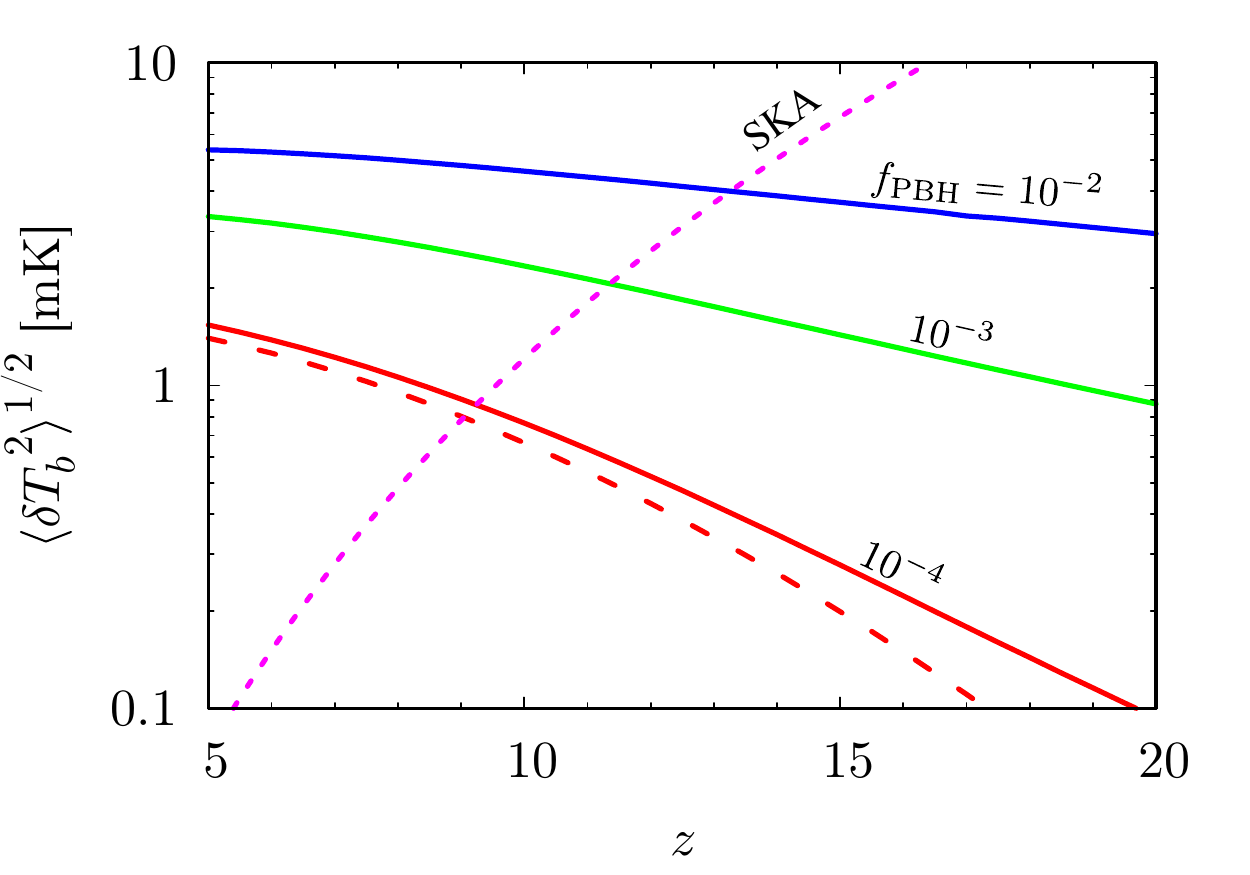}
\label{subfig:deltaT_b_4}
}
\caption{
The 3$\sigma$ fluctuation of the differential brightness temperature of redshifted 21cm emissions. Different coloured solid lines in each panel correspond to different values of $f_\pbh$. The dashed red lines and dotted magenta lines respectively represent the conventional prediction and the noise curve of SKA.
}
\label{fig:deltaT_b}
\end{figure}

In Figure~\ref{fig:deltaT_b}, we show the 3$\sigma$ fluctuation of the differential brightness temperature \eqref{eq:rms-Tb} in the presence of the minihaloes from the PBH Poisson fluctuations. The solid lines correspond to the cases with non-zero $f_\pbh$, while the red dashed line represents the prediction from the standard scenario with adiabatic Gaussian primordial perturbations and the dotted magenta line expresses the noise curve of SKA-like surveys given by \cite{Furlanetto:2006jb}
\begin{equation}
\delta T_{\rm noise} = 20~{\rm mK}~\frac{10^4~{\rm m}^2}{A_{\rm tot}} \bigg( \frac{10~{\rm arcmin}}{\Delta\theta_{\rm beam}} \bigg)^2 
\bigg( \frac{1+z}{10} \bigg)^{4.6} \bigg( \frac{{\rm MHz}}{\Delta\nu_{\rm band}} \frac{100h}{t_{\rm int}}\bigg)^{1/2} \, ,
\end{equation}
where $A_{\rm tot}$ is the effective collecting area of the radio arrays and $t_{\rm int}$ is the integration time. We have set $A_{\rm tot} = 10^5~{\rm m}^2$, $\Delta\theta_{\rm beam} = 9$ arcmin, $\Delta\nu_{\rm band} = 1$ MHz and $t_{\rm int} = 1000$ h. As we can see, for example, PBHs with $M_\pbh = 10 M_\odot$ can significantly alter the 21cm signal for $f_\pbh \gtrsim 10^{-3}$. This indicates that future 21cm observations can place strong constraints on the abundance of PBHs.

\begin{figure}[tp]
\centering
\includegraphics [width = 12cm, clip]{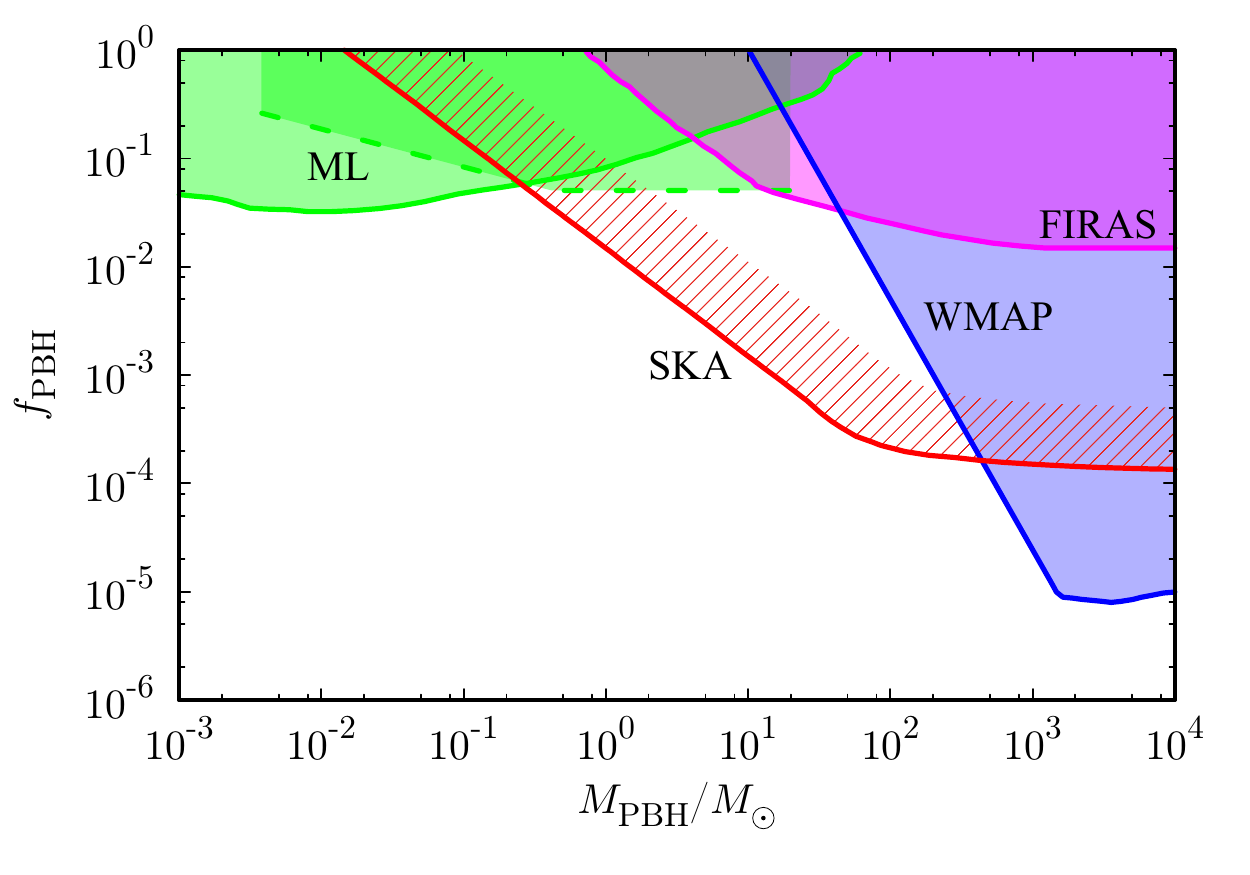}
\caption{
Forecast constraint on the PBH dark matter fraction in terms of the PBH mass by SKA (inside the red hatched region). Other shaded regions represent the current constraints summarized in \cite{Carr:2016drx}.
}
\label{fig:PBHconstraint}
\end{figure}

As a summary, we show in Figure~\ref{fig:PBHconstraint} the forecast constraint on the PBH fraction by SKA using the $\chi^2$ analysis following \cite{Sekiguchi:2013lma}. Inside the red hatched region, SKA can observe the 21cm signals with 95\% confidence level.

\section{Conclusions} 
\label{sec:conc}

In this article, we have studied the impacts of PBHs as a part of cold dark matter on the early small-scale structure. Since the formation of PBHs is a rare event and thus follows Poisson distribution, there exists an additional scale-invariant fluctuation due to the Poisson noise of the number of PBHs, which is independent of the primordial adiabatic density perturbation. Thus, it behaves as an isocurvature perturbation and, since it rapidly grows on small scales, early structure formation is accelerated. We have found that the halo mass function is significantly enhanced due to the addition of the constant power spectrum \eqref{eq:pbh_powerspec} to the variance \eqref{eq:smoothedvariance} as shown in Figure~\ref{fig:mf1}. Those haloes originated from the PBH Poisson fluctuations can further enhance the 21cm emission line signals by providing modified mass function. The enhanced fluctuation of the 21cm differential brightness temperature shown in Figure~\ref{fig:deltaT_b} is very sensitive to the number of minihaloes seeded by PBHs, even if PBHs occupy sub-percent fraction of the present cold dark matter. Thus precise future radio telescope surveys such as SKA can provide very strong constraints on the abundance of PBHs. 
In this article, we have neglected the effect of matter accretion onto PBHs and its X-ray emission on the 21cm fluctuations discussed in \cite{Tashiro:2012qe}. Such an extra X-ray heating can also alter the 21cm brightness temperature significantly and it can strongly constrain the PBH abundance. Moreover, the enhancement of small-scale structure at high redshift can modify the reionization history. In particular, the enhancement of halo formation leads to the increase of luminous sources which emits UV photons ionizing IGM. Hence the reionization optical depth is altered and it can be constrained by the CMB observation as discussed in \cite{Sekiguchi:2013lma} in the case of our interest. We leave the combined analysis with the above issues in future work.

\subsection*{Acknowledgment}

We acknowledge the support from the Korea Ministry of Education, Science and Technology, Gyeongsangbuk-Do and Pohang City for Independent Junior Research Groups at the Asia Pacific Center for Theoretical Physics.
JG is also supported in part by a TJ Park Science Fellowship of POSCO TJ Park Foundation, and by the Basic Science Research Program through the National Research Foundation of Korea (NRF) Research Grant NRF-2016R1D1A1B03930408. 
NK acknowledges the support by Grant-in-Aid for JSPS Fellows.

\providecommand{\href}[2]{#2}\begingroup\raggedright\endgroup


\end{document}